\documentclass[amsmath,amssymb,twocolumn]{revtex4}
\usepackage{graphicx}% Include figure files
\usepackage{bm}% bold math
\makeatletter \@addtoreset{equation}{section} \makeatother
\renewcommand{\theequation}{\arabic{section}.\arabic{equation}}

\begin{document}
\title{Rogue waves in the Davey-Stewartson equation}
\author{Yasuhiro Ohta$^{1}$\footnote{Email: ohta@math.kobe-u.ac.jp}
\hspace{0.15cm}
and \hspace{0.1cm}  Jianke Yang$^{2}$\footnote{Email: jyang@math.uvm.edu}}
\affiliation{
{\small\it $^1$Department of Mathematics, Kobe University, Rokko,
Kobe 657-8501, Japan} \\ {\small\it $^2$Department of Mathematics
and Statistics, University of Vermont, Burlington, VT $05401$,
U.S.A}}

\begin{abstract}
General rogue waves in the Davey-Stewartson-I equation are derived
by the bilinear method. It is shown that the simplest (fundamental)
rogue waves are line rogue waves which arise from the constant
background with a line profile and then disappear into the constant
background again. It is also shown that multi-rogue waves describe
the interaction of several fundamental rogue waves. These
multi-rogue waves also arise from the constant background and then
decay back to it, but in the intermediate times, interesting curvy
wave patterns appear. However, higher-order rogue waves are found to show more interesting
features. Specifically, only part of the wave structure in the
higher-order rogue waves rises from the constant background and then
retreats back to it, and this transient wave exhibits novel
patterns such as parabolas. But the other part of the wave
structure comes from the far distance as a localized lump, which decelerates to the near field and interacts
with the transient rogue wave, and is then
reflected back and accelerates to the large distance again. These rogue-wave solutions have
interesting implications for two-dimensional surface water waves in
the ocean.
\end{abstract}

%\pacs{05.45.Yv}
%05.45.Yv: Solitons

\maketitle

\section{Introduction}
Rogue waves are large and spontaneous ocean surface waves that occur
in the sea and are a threat even to large ships and ocean liners
\cite{rogue_water}. Recently, an optical analogue of rogue waves ---
optical rogue waves, was also observed in optical fibres
\cite{Rogue_nature1,Rogue_nature2}. A growing consensus is that both
oceanic and optical rogue waves appear as a result of modulation
instability of monochromatic nonlinear waves. Mathematically, the
first and simplest rogue-wave solution was reported in the nonlinear
Schr\"odinger (NLS) equation by Peregrine \cite{Peregrine}. This
solution approaches a non-zero constant background as time goes to
$\pm \infty$ but develops a localized hump with peak amplitude three times the constant
background in the intermediate times. Recently, higher-order rogue
waves in the NLS equation were reported in many articles
\cite{Akhmediev_PRE,Rogue_higher_order,Rogue_higher_order2,Rogue_Gaillard,Rogue_triplet,Rogue_circular,Liu_qingping,
OY}. It was shown that these higher-order waves could reach higher
peak amplitudes or exhibit multiple intensity peaks at different
spatial locations and times. In addition to the NLS equation, rogue
waves in some other wave equations (such as the Hirota equation)
have also been explored \cite{rogue_Hirota}. Rogue waves are
intimately related to homoclinic solutions which approach a constant
background as time goes to $\pm \infty$ but develop
spatially-periodic wave patterns in the intermediate times
\cite{Akhmediev_1985,Akhmediev_1988,Its_1988,Ablowitz_homo}. Indeed,
rogue waves can be obtained from homoclinic solutions when the
spatial period of homoclinic solutions goes to infinity
\cite{Rogue_Gaillard,Akhmediev_1985,Akhmediev_1988,Rogue_homo}.

Rogue waves which have been studied so far are mostly
one-dimensional. But ocean surface waves are always two-dimensional.
Thus a natural question is to investigate rogue waves in
two-dimensional model equations. It is well
known that the evolution of a two-dimensional wave-packet on water
of finite depth is governed by the Benney-Roskes-Davey-Stewartson
equation \cite{Benney_Roskes,Davey_Stewartson,Ablowitz_book}. This
equation is often just called the Davey-Stewartson (DS) equation in
the literature. The DS equation is divided into two types, DSI and
DSII equations, depending on the signs of its coefficients. The
simplest (one-mode) homoclinic solution to the DS equation was
derived in \cite{Tajiri_1999}. Taking the spatial period of this
homoclinic solution to go to infinity, the simplest (fundamental)
rogue-wave solution was also obtained there. But more general rogue
waves in the DS equation are still unknown.

In this paper, general rogue waves in the DS-I equation are derived.
These solutions are obtained by the bilinear method and expressed in
terms of determinants. It is shown that the simplest (fundamental)
rogue waves are line rogue waves which arise from the constant
background with a line profile and then disappear into the constant
background again (this simplest rogue wave agrees with that reported
in \cite{Tajiri_1999}). It is also shown that non-fundamental rogue waves contain different types such as the
multi-rogue waves and higher-order rogue waves. The multi-rogue waves
describe the interaction of several fundamental rogue waves. These
multi-rogue waves arise from the constant background and then decay
back to it, but in the intermediate times, interesting curvy wave
patterns appear. Higher-order rogue waves, on the other hand, exhibit certain features
which are very novel. Specifically, only parts of the wave
structures in these higher-order rogue waves rise from the constant
background and then retreat back to it, exhibiting unusual transient wave
patterns (such as parabola shapes) in the intermediate times. But
the other parts of the waves come from the far distance as localized lumps,
which interact with the transient rogue waves in the near field and then
are reflected back to the large distance again. Since the DS equation is a
well known mathematical model for two-dimensional surface water
waves, these general rogue-wave solutions can have interesting
implications for the study of rogue waves in the ocean.

\section{Rational solutions in the Davey-Stewartson-I equation} \label{ratsol}

The Davey-Stewartson-I (DSI) equation is given by
\begin{equation} \label{DSI}
\begin{array}{l}
iA_t=A_{xx}+A_{yy}+(\epsilon |A|^2-2Q)A,
\\[5pt]
Q_{xx}-Q_{yy}=\epsilon(|A|^2)_{xx},
\end{array}
\end{equation}
where $\epsilon=1$ or $-1$. It is noted that under the variable
transformation $Q\to Q+\epsilon |A|^2$, $x\leftrightarrow y$ and
$\epsilon\to -\epsilon$, this equation is invariant,
thus we can fix the sign of $\epsilon$ without loss of generality.
However the transformation $Q\to Q+\epsilon |A|^2$
changes the boundary condition of $Q$ in general, thus we
keep $\epsilon$ in our analysis. Equation (\ref{DSI}) is transformed
into the bilinear form,
\begin{equation} \label{bilinDSI}
\begin{array}{l}
(D_x^2+D_y^2-iD_t)g\cdot f=0,
\\[5pt]
(D_x^2-D_y^2)f\cdot f=2\epsilon(f^2-|g|^2),
\end{array}
\end{equation}
through the variable transformation,
\begin{equation} \label{vartrans}
A=\sqrt{2}\frac{g}{f},
\quad
Q=\epsilon-(2\log f)_{xx},
\end{equation}
where $f$ is a real variable and $g$ is a complex one.

Rogue waves are rational solutions (under certain parameter
restrictions). Thus we first present the general rational solutions
to the DSI equation in the following theorem. The proof of this theorem
is given in the appendix.

\vspace{0.3cm}
\noindent{\bf Theorem 1} \ The DSI equation (\ref{DSI}) has rational
solutions (\ref{vartrans}) with $f$ and $g$ given by $N\times N$
determinants
\begin{equation} \label{formula_rational}
f=\tau_0,
\quad
g=\tau_1,
\end{equation}
where $\tau_n=\det_{1\le i,j\le N}\left(m_{ij}^{(n)}\right)$, and the matrix
elements are given by either

\noindent (a)
\begin{eqnarray}
\hspace{-2cm} m_{ij}^{(n)} & = & \sum_{k=0}^{n_i}c_{ik}
(p_i\partial_{p_i}+\xi'_i+n)^{n_i-k}   \nonumber \\
&& \sum_{l=0}^{n_j}\bar c_{jl}
(\bar p_j\partial_{\bar p_j}+\bar\xi'_j-n)^{n_j-l}
\frac{1}{p_i+\bar p_j},
\end{eqnarray}
\begin{equation} \label{xiprime}
\xi'_i=\frac{p_i-\epsilon p_i^{-1}}{2}x+\frac{p_i+\epsilon p_i^{-1}}{2}y
+\frac{p_i^2+p_i^{-2}}{\sqrt{-1}}t,
\end{equation}

\noindent
or

\noindent (b)
\begin{equation}
m_{ij}^{(n)}=\sum_{\nu=0}^{n_i+n_j}
\left(\frac{-1}{p_i+\bar p_j}\right)^{\nu+1}
(\partial_x+\partial_y)^\nu P_i^{(n)}\overline{P_j^{(-n)}},
\end{equation}
\begin{equation}
P_i^{(n)}=\sum_{k=0}^{n_i}\hat c_{ik}S_{n_i-k}
\left(\mbox{\boldmath $\xi$}^{(n)}(p_i)\right),
\end{equation}
where $S_n(\mbox{\boldmath $x$})$ is the elementary Schur polynomial
defined via the generating function
\[
\sum_{n=0}^{\infty}S_n(\mbox{\boldmath $x$})\lambda^n
=\exp\left(\sum_{k=1}^{\infty}x_k\lambda^k\right)
\]
for $\mbox{\boldmath $x$}=(x_1,x_2,\cdots)$,
\[
\mbox{\boldmath $\xi$}^{(n)}(p)
=[\xi_1(p)+n,\;  \xi_2(p),  \, \cdots, \, \xi_k(p)+\delta_{k1}n, \, \cdots],
\]
$\delta_{ij}$ is the Kronecker delta notation (which is equal to 1 when $i=j$ and zero otherwise), and
\begin{eqnarray*}
\xi_k(p) & = & \frac{1}{k!}
\left[\frac{p+\epsilon (-1)^k/p}{2}x+\frac{p-\epsilon (-1)^k/p}{2}y  \right.   \\
&& \hspace{0.6cm} \left. +\frac{2^kp^2-(-2)^k/p^2}{2\sqrt{-1}}t\right].
\end{eqnarray*}
In (a) and (b),
$p_i$, $c_{ik}$ and $\hat c_{ik}$ are arbitrary complex constants, and $n_i$ is an
arbitrary positive integer. These two expressions in (a) and (b) would yield identical solutions if the
constants $c_{ik}$ and $\hat c_{ik}$ in them are related by
\[\hat c_{ik}=c_{ik}(n_i-k)!, \quad \hat d_{jl}=d_{jl}(n_j-l)!.\]
Thus in the later text we use the expression in (a).
By a scaling of $f$ and $g$, we can normalize $c_{i0}=1$ without
loss of generality, thus hereafter we set $c_{i0}=1$. We will also call
the above solution as the $N$-rational solution of order $(n_1,n_2,\cdots,n_N)$.
We comment that a more explicit expression for $m_{ij}^{(n)}$ similar
to (2.6) in \cite{OY} can also be obtained, but since that expression is a bit
complicated, we omit it in this paper.

The simplest rational solution, namely 1-rational solution of 1st
order, is given by taking $N=1$ and $n_1=1$,
\begin{eqnarray*}
&& \hspace{-0.35cm}  f=\sum_{k=0}^1c_{1k}
(p_1\partial_{p_1}+\xi'_1)^{1-k}
\sum_{l=0}^1\bar c_{1l}
(\bar p_1\partial_{\bar p_1}+\bar\xi'_1)^{1-l}
\frac{1}{p_1+\bar p_1}
\\
&& \hspace{-0.35cm}
=\left(p_1\partial_{p_1}+\xi'_1+c_{11}\right)
\left(\bar p_1\partial_{\bar p_1}+\bar\xi'_1+\bar c_{11}\right)
\frac{1}{p_1+\bar p_1}
\\
&& \hspace{-0.35cm}
=\frac{1}{p_1+\bar p_1}\left[
\left(\xi'_1+c_{11}-\frac{p_1}{p_1+\bar p_1}\right)
\left(\bar\xi'_1+\bar c_{11}-\frac{\bar p_1}{p_1+\bar p_1}\right) \right. \\
&& \left. \hspace{1.5cm} +\frac{p_1\bar p_1}{(p_1+\bar p_1)^2}\right],
\end{eqnarray*}
\begin{eqnarray*}
&&\hspace{-0.35cm}  g=\sum_{k=0}^1c_{1k}
(p_1\partial_{p_1}+\xi'_1+1)^{1-k}   \\
&& \hspace{0.3cm} \sum_{l=0}^1\bar c_{1l}
(\bar p_1\partial_{\bar p_1}+\bar\xi'_1-1)^{1-l}
\frac{1}{p_1+\bar p_1}
\\
&& \hspace{-0.35cm}
=\left(p_1\partial_{p_1}+\xi'_1+1+c_{11}\right)
\left(\bar p_1\partial_{\bar p_1}+\bar\xi'_1-1+\bar c_{11}\right)
\frac{1}{p_1+\bar p_1}
\\
&& \hspace{-0.35cm}
=\frac{1}{p_1+\bar p_1}\left[
\left(\xi'_1+1+c_{11}-\frac{p_1}{p_1+\bar p_1}\right) \times \right. \\
&&
\hspace{1.5cm} \left. \left(\bar\xi'_1-1+\bar c_{11}-\frac{\bar p_1}{p_1+\bar p_1}\right)
+\frac{p_1\bar p_1}{(p_1+\bar p_1)^2}\right],
\end{eqnarray*}
where
\[
\xi'_1=\frac{p_1-\epsilon p_1^{-1}}{2}x+\frac{p_1+\epsilon p_1^{-1}}{2}y
+\frac{p_1^2+p_1^{-2}}{i}t,
\]
and $p_1$, $c_{11}$ are arbitrary complex constants. This
solution can be rewritten as
\begin{eqnarray*}
f=\frac{1}{p_1+\bar p_1}(\xi\bar\xi+\Delta), \hspace{0.3cm}
g=\frac{1}{p_1+\bar p_1}[(\xi+1)(\bar\xi-1)+\Delta],
\end{eqnarray*}
where
\[
\xi=ax+by+\omega t+\theta, \quad \Delta =p_1\bar p_1/(p_1+\bar
p_1)^2,
\]
and
\[
a=(p_1-\epsilon p_1^{-1})/2, \quad b=(p_1+\epsilon p_1^{-1})/2,
\]
\[
\omega=(p_1^2+p_1^{-2})/i, \quad \theta=c_{11}-p_1/(p_1+\bar
p_1).
\]
If we separate the real and imaginary parts of $a, b, \omega$ and
$\theta$ as
\[
a=a_1+ia_2, \hspace{0.2cm} b=b_1+ib_2, \hspace{0.2cm} \omega=\omega_1+i\omega_2, \hspace{0.2cm}
\theta=\theta_1+i\theta_2,
\]
then
\begin{equation*}
A(x,y,t)=\sqrt{2}\left[1-\frac{2i(a_2x+b_2y+\omega_2t+\theta_2)+1}{W}\right],
\end{equation*}
where
\[W=(a_1x+b_1y+\omega_1t+\theta_1)^2+(a_2x+b_2y+\omega_2t+\theta_2)^2+\Delta.\]
The solution $Q$ can also be written down from (\ref{vartrans}) and
the above $f$.

This simplest rational solution has three distinctly different
dynamical behaviors depending on the parameter value of $p_1^2$.
\begin{enumerate}
\item If $p_1^2$ is not real, then it is easy to see that $b/a$ is not
real, hence $b_1/b_2\ne a_1/a_2$. In this case, along the
$[x(t),y(t)]$ trajectory where
\[a_1x+b_1y=-\omega_1t, \quad a_2x+b_2y=-\omega_2t, \]
$(A, Q)$ are constants. In addition, at any given time, $(A, Q) \to
(\sqrt{2}, 1)$ when $(x,y)$ goes to infinity. Thus the solution is a
localized soliton moving on a constant background.

\item If $p_1^2<0$, i.e., $p_1$ is purely imaginary, then $a, b$ and $\omega$ are all imaginary.
In this case, the solution is a function of $a_2x+b_2y+\omega_2t$
only, and is thus a line soliton moving on a constant background.

\item If $p_1^2>0$, i.e., $p_1$ is real, then $a, b$ are real
but $\omega$ is imaginary. In this case, the solution is also a line
wave, but it is not a moving line soliton anymore. As $t\to \pm
\infty$, this line wave goes to a uniform constant background; in
the intermediate times, it rises to a higher amplitude. Thus this is
a line wave which ``appears from nowhere and disappears with no
trace", hence it is a line rogue wave.
\end{enumerate}

{}From the above analysis, we see that rational solutions
(\ref{formula_rational}) to the DSI equation become rogue waves when
the parameters $p_i$ are real-valued (this fact holds for $N=1$ as
well as for higher $N$ integers). In the next section, we will
examine these rogue waves in more detail.

It is noted from the above explicit solution formulae that the
parameter $c_{11}$ causes a shift of the origin in the $(x,y,t)$
space. Thus $c_{11}$ can be set to zero by a shift of the $(x,y,t)$
axes. This fact also holds for $N=1$ as well as for higher $N$
integers in the solution (\ref{formula_rational}).

\section{Rogue waves in the Davey-Stewartson-I equation}

As we have shown above, rational solutions (\ref{formula_rational})
in Theorem 1 become rogue waves in the DSI equation when all
parameters $p_i$ are required to be real. In this section, we
analyze the dynamics of these rogue waves in detail.

\subsection{Fundamental rogue waves}
\begin{figure}[h!]
\centerline{\includegraphics[width=0.5\textwidth]{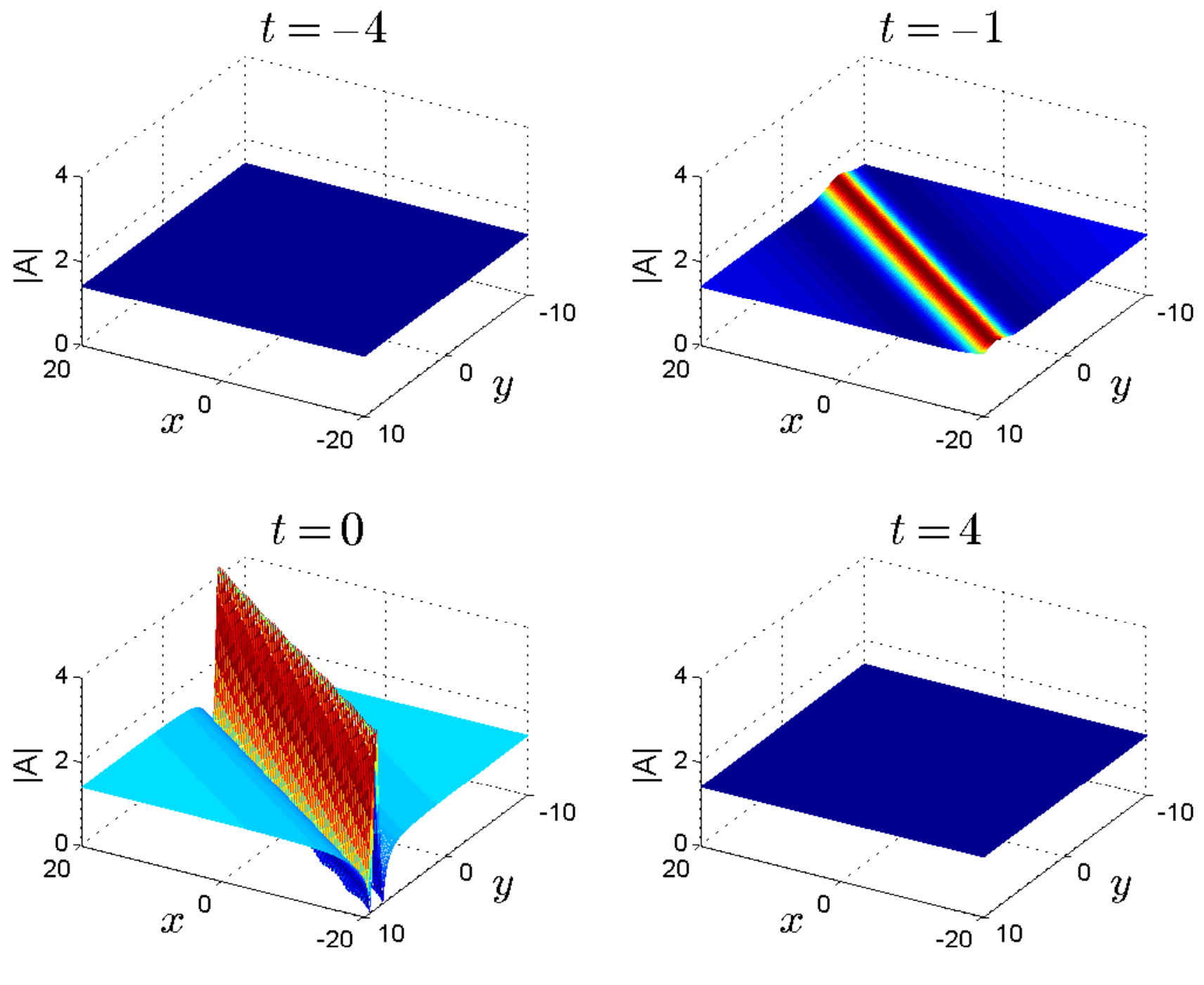}}
\caption{A fundamental rogue wave (\ref{formula_line_rogue}) in the
DSI equation with parameters $\epsilon=1, p_1=1.5$. }
\label{f:line_rogue}
\end{figure}

The fundamental rogue waves in the DSI equation are obtained when
one takes $N=1$, $n_1=1$ and $p_1$ real in the rational solution
(\ref{formula_rational}), and $c_{11}$ is a free complex parameter.
After a shift of time and space coordinates, $c_{11}$ can be
eliminated and the fundamental rogue waves can be written as
\begin{equation} \label{formula_line_rogue}
A(x,y,t)=\sqrt{2}\left[1+\frac{8i\Omega
t-4}{1+(k_1x+k_2y)^2+4\Omega^2t^2}\right],
\end{equation}
\begin{equation}
Q(x,y,t)=1-4\epsilon k_1^2
\frac{1-(k_1x+k_2y)^2+4\Omega^2t^2}{[1+(k_1x+k_2y)^2+4\Omega^2t^2]^2},
\end{equation}
where
\[k_1=p_1-\epsilon p_1^{-1}, \quad  k_2=p_1+\epsilon p_1^{-1}, \quad
\Omega=p_1^2+p_1^{-2}.
\]
This solution describes a line rogue wave with the line oriented in
the $(k_1, k_2)$ direction of the $(x,y)$ plane, thus the
fundamental rogue waves in the DSI equation are line rogue waves.
Along the line direction (with $k_1x+k_2y$ fixed), the solution is a
constant. As $t\to \pm \infty$, the solution $A$ uniformly
approaches the constant background $\sqrt{2}$ everywhere in the
$(x,y)$ plane; but in the intermediate times, $|A|$ reaches maximum
amplitude $3\sqrt{2}$ (i.e., three times the background amplitude)
on the line $k_1x+k_2y=0$ at time $t=0$. This fundamental rogue wave
is illustrated in Fig. \ref{f:line_rogue} with parameters
$\epsilon=1$ and $p_1=1.5$.

The above fundamental rogue waves in the DSI equation are
two-dimensional counterparts of the fundamental (Peregrine) rogue
waves in the NLS equation \cite{Peregrine,Akhmediev_PRE}. Indeed,
when we take $\epsilon=1$ and $p_1=1$ in the above fundamental rogue
waves, we have $k_1=0$, hence the solution $A$ is independent of $x$
and $Q=1$. In this case, the DSI equation reduces to the NLS
equation, and this fundamental rogue wave of the DSI equation
reduces to the Peregrine rogue wave of the NLS equation.

Non-fundamental rogue waves can be obtained from the $N$-rational
solutions of order $(n_1,n_2,\cdots,n_N)$ in Eq.
(\ref{formula_rational}) with real values of $(p_1, \dots, p_N)$
when $N>1$, or $n_1>1$, or both. Below we consider two subclasses of
these non-fundamental rogue waves.

\subsection{Multi-rogue waves}

One subclass of non-fundamental rogue waves is the multi-rogue
waves, which are obtained when we take $N>1$, $n_1=\dots=n_N=1$ in
the rational solution (\ref{formula_rational}) with real values of
$(p_1, \dots, p_N)$. These rogue waves describe the interaction of
$N$ individual fundamental rogue waves. When $t\to \pm \infty$, the
solution approaches the constant background uniformly in the entire
$(x,y)$ plane. In the intermediate times, $N$ line rogue waves arise
from the constant background, interact with each other, and then
disappear into the background again. In the far field of the $(x,y)$
plane, the solution consists of $N$ separate line rogue waves.
However, in the near field where these line rogue waves intersect
and interact, wavefronts of the solution are no longer lines, and
interesting curvy wave patterns would appear.

\begin{figure}[h!]
\centerline{\includegraphics[width=0.5\textwidth]{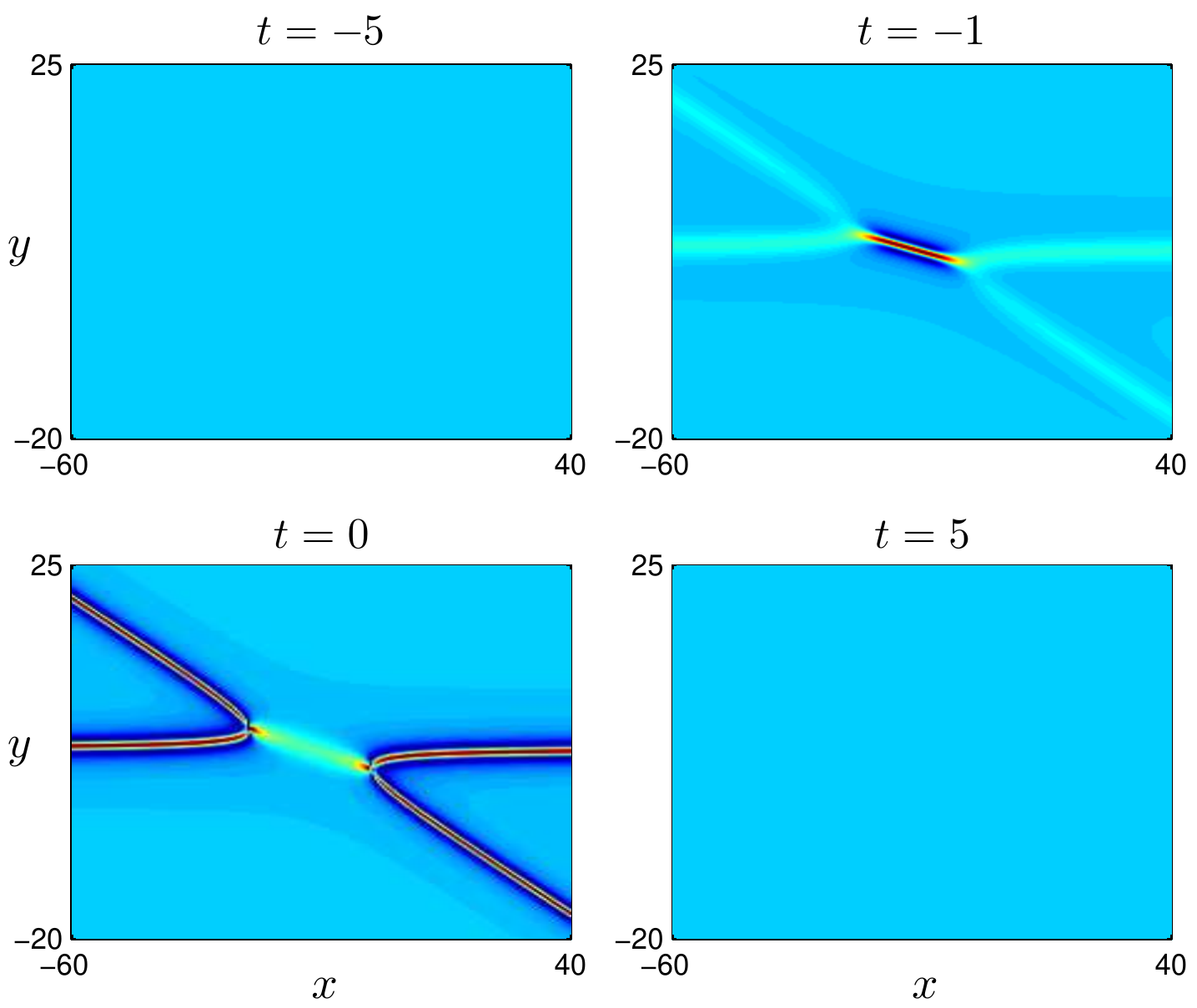}}
\caption{A two-rogue-wave solution (\ref{s:fg2rogue}) in the DSI
equation for parameters (\ref{para_2line_rogue}). Plotted is the
$|A|$ field. The constant-background value is $\sqrt{2}$. }
\label{f:2line_rogue}
\end{figure}

To demonstrate these multi-rogue-wave solutions, we first consider
the $N=2$ case. In this case, the $f$ and $g$ functions of the
solutions can be obtained from (\ref{formula_rational}) as
\begin{equation}\label{s:fg2rogue}
f=\left|\begin{matrix} m_{11}^0 &m_{12}^0 \cr m_{21}^0 &m_{22}^0 \end{matrix} \right|,
\quad g=\left|\begin{matrix} m_{11}^1 &m_{12}^1 \cr m_{21}^1
&m_{22}^1 \end{matrix} \right|,
\end{equation}
where
\begin{eqnarray*}
m_{ij}^0 & = & \frac{1}{p_i+\bar p_j}\left[
\left(\xi'_i+c_{i1}-\frac{p_i}{p_i+\bar p_j}\right)\times  \right. \\
&& \hspace{1.4cm} \left. \left(\bar\xi'_j+\bar c_{j1}-\frac{\bar p_j}{p_i+\bar p_j}\right)
+\frac{p_i\bar p_j}{(p_i+\bar p_j)^2}\right],
\end{eqnarray*}
\begin{eqnarray*}
m_{ij}^1 & = & \frac{1}{p_i+\bar p_j}\left[
\left(\xi'_i+1+c_{i1}-\frac{p_i}{p_i+\bar p_j}\right) \times  \right. \\
&& \hspace{1.2cm} \left. \left(\bar\xi'_j-1+\bar c_{j1}-\frac{\bar p_j}{p_i+\bar p_j}\right)
+\frac{p_i\bar p_j}{(p_i+\bar p_j)^2}\right],
\end{eqnarray*}
$\xi'_j$ is given by Eq. (\ref{xiprime}), $p_1, p_2$ are free real
parameters, and $c_{11}$, $c_{21}$ are free complex parameters. The
complex parameter $c_{11}$ can be removed by a shift of the
$(x,y,t)$ axes, then this two-rogue-wave solution contains four
non-trivial real parameters, namely, $p_1$, $p_2$, and the real and
imaginary parts of $c_{21}$. This solution for parameters
\begin{equation} \label{para_2line_rogue}
\epsilon=1, \ p_1=1, \ p_2=1.5, \  c_{11}=c_{21}=0
\end{equation}
is shown in Fig. \ref{f:2line_rogue}. It is seen that when these two
line rogue waves arise from the constant background, the region of
their intersection acquires higher amplitude first (see $t=-1$
panel). After these higher amplitudes in the intersection region
fade, the line-rogue solutions in the far field then rise to higher
amplitude (see $t=0$ panel). Interestingly, the wave pattern at
$t=0$ features two curvy wavefronts which are well separated. These
curvy wavefronts are caused by the interaction of the two
fundamental (line) rogue waves. At large times, the solution goes
back to the constant background again (see $t=5$ panel).

\begin{figure}[h!]
\centerline{\includegraphics[width=0.5\textwidth]{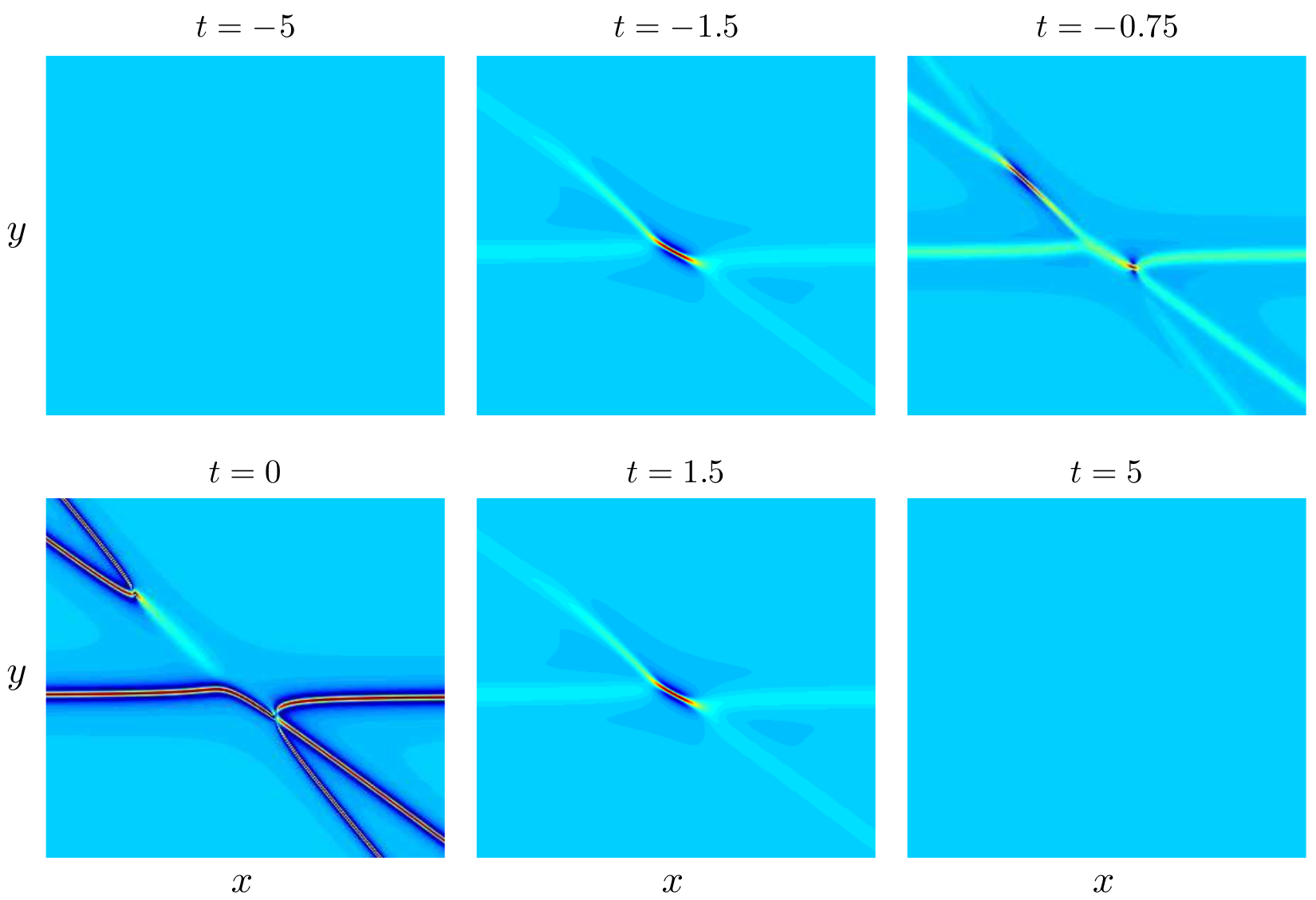}}
\caption{A three-rogue-wave solution in the DSI equation for
parameters (\ref{para_3line_rogue}). Plotted is the $|A|$ field.}
\label{f:3line_rogue}
\end{figure}

For larger $N$, these multi-rogue waves have qualitatively similar
behaviors, except that more line-rogue waves will arise and interact
with each other, and more complicated wavefronts will form in the
interaction region. For example, with $N=3$ and parameter choices
\begin{equation} \label{para_3line_rogue}
\epsilon=1, \ p_1=1, \ p_2=1.5, \ p_3=2,  \ c_{11}=c_{21}=c_{31}=0,
\end{equation}
the corresponding solution is shown in Fig. \ref{f:3line_rogue}. As
can be seen, the transient solution patterns become more intricate.

\subsection{Higher-order rogue waves}

Another subclass of non-fundamental rogue waves is the higher-order
rogue waves, which are obtained when we take $N=1$ and $n_1>1$ in
the rational solution (\ref{formula_rational}) with a real value of
$p_1$. For instance, if $n_1=2$, we get second-order rogue waves
from (\ref{formula_rational}) as
\begin{eqnarray}  \label{f:2nd_order}
f & = & \left[(p_1\partial_{p_1}+\xi'_1)^2 +c_{12}\right] \nonumber \\
&& \left[(\bar p_1\partial_{\bar p_1}+\bar\xi'_1)^2 +\bar c_{12}\right]
\frac{1}{p_1+\bar p_1},
\end{eqnarray}
\begin{eqnarray}  \label{g:2nd_order}
g & = & \left[(p_1\partial_{p_1}+\xi'_1+1)^2 +c_{12}\right ]   \nonumber \\
&& \left[(\bar
p_1\partial_{\bar p_1}+\bar\xi'_1-1)^2 +\bar c_{12}\right]
\frac{1}{p_1+\bar p_1},
\end{eqnarray}
where $\xi'_1$ is given by Eq. (\ref{xiprime}), $p_1$ is a free real
parameter, and $c_{12}$ is a free complex parameter. Here we have
set $c_{11}=0$ in (\ref{formula_rational}) by a shift of the
$(x,y,t)$ axes. Higher-order rogue waves with $n_1>2$ can be
similarly obtained.

\begin{figure}[h!]
\centerline{\includegraphics[width=0.5\textwidth]{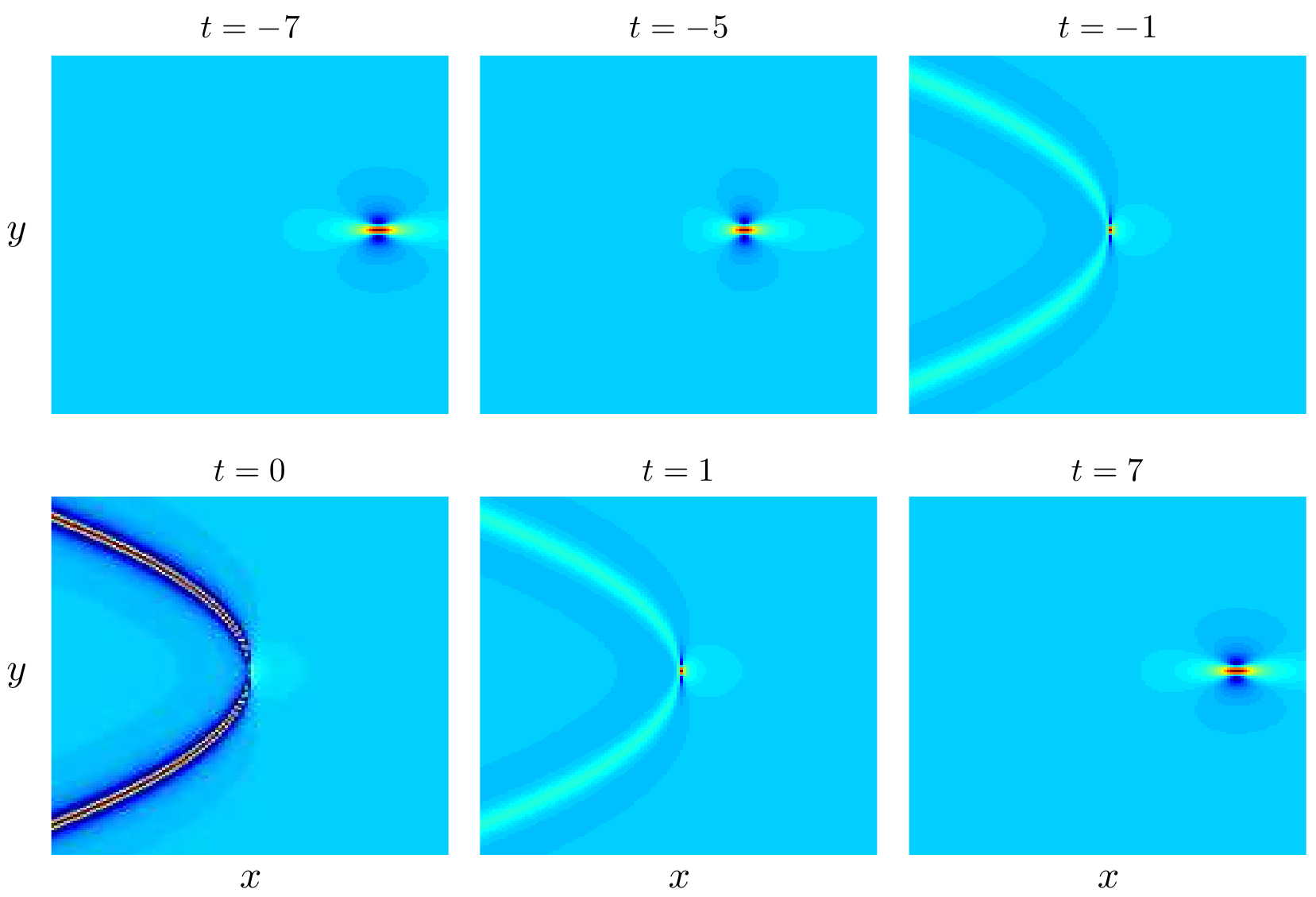}}
\caption{A second-order rogue wave solution (\ref{A_second_order})
in the DSI equation with $\epsilon=1$. }
\label{f:higher_order_rogue}
\end{figure}

An interesting phenomenon is that, unlike the multi-rogue waves
discussed in the previous subsection, these higher-order rogue waves
do \emph{not} uniformly approach the constant background as $t\to
\pm \infty$. Instead, only parts of their wave structures approach
the constant background as $t\to \pm \infty$, but the other parts
move to the far distance as localized lumps with undiminished amplitude
and increasing velocity as $t\to \pm \infty$. To illustrate these
behaviors, we consider the above second-order rogue waves. For
parameter values
\begin{equation} \label{para_2nd_order_rogue}
\epsilon=1, \quad p_1=1, \quad c_{12}=0,
\end{equation}
we get
\begin{equation*}
f=\frac{1}{2}\left[x-(4t^2+y-y^2)\right]^2+\left[(y-\frac{1}{2})^2+\frac{1}{4}\right](8t^2+\frac{1}{2}),
\end{equation*}
\begin{equation*}
g=f-4it(x+y-y^2-4t^2)+x+y-y^2-12t^2-\frac{1}{2},
\end{equation*}
and
\begin{equation} \label{A_second_order}
A(x,y,t)=\sqrt{2}\hspace{0.1cm} \frac{g}{f}.
\end{equation}
For this solution, $\bar{A}(x,y,t)=A(x,y,-t)$, thus
$|A(x,y,-t)|=|A(x,y,t)|$. This solution is displayed in Fig.
\ref{f:higher_order_rogue}. We see that when $|t|\gg 1$, the solution
is a localized lump sitting on the constant background $\sqrt{2}$ (see $t=\pm 7$ panels).
The peak amplitude of the lump is attained at
\[
(x, y)=(4t^2+\frac{1}{4}, \ \frac{1}{2}),
\]
thus this lump is accelerating rightward as $|t|$ increases (see $t=-7$ and $t=-5$
panels). The peak-amplitude value of the lump stays at $3\sqrt{2}$ and is
unchanged though. As the lump accelerates rightward (with increasing $|t|$),
its vertical size (along the $y$ direction)
remains the same, but its horizontal size (along the $x$ direction)
expands (see $t=-7, -5, -1$ panels) . When $t\to 0$, this lump disappears. At the same time, a
parabola-shaped rogue wave rises from the background (see $t=-1$ and $t=0$
panels). At $t=0$, this parabola is located at
\[x=y-y^2, \]
where the rogue wave reaches peak amplitude $3\sqrt{2}$
(see $t=0$ panel). Visually one may describe the solution in Fig.
\ref{f:higher_order_rogue} as an incoming lump being reflected back
by the emergence of a parabola-shaped rogue wave. In addition, this lump decelerates
as it comes afar and accelerates as it goes away.

We have also examined the second-order rogue waves
(\ref{f:2nd_order})-(\ref{g:2nd_order}) for other parameter choices
of $(\epsilon, p_1, c_{12})$, and found that those solutions are
qualitatively the same as the one in Fig. 4, except that those
solution patterns may be stretched and skewed in the $(x,y)$ plane.

\section{Summary}
In summary, we have derived general rogue waves in the DSI equation
by the bilinear method, and our solutions are given in terms of
determinants. We showed that the simplest (fundamental) rogue waves
are line rogue waves which arise from the constant background with a
line profile and then disappear into the constant background again
(see Fig. 1). We also showed that multi-rogue waves describe the
interaction of several fundamental rogue waves, and interesting
curvy wave patterns appear due to this interaction (see Figs. 2 and
3). However, higher-order rogue waves were found to show very novel
features. Specifically, only parts of the wave structures in the
higher-order rogue waves rise from the constant background and then
retreat back to it, but the other parts of the waves come from the far distance
as localized lumps, which interact with the transient
rogue waves in the near field and then are reflected back and accelerate to the large distance
again (see Fig. 4). These rogue wave solutions to the DSI
equation generalize the rogue waves of the NLS equation into two
spatial dimensions, and they could play a role in the physical
understanding of rogue water waves in the ocean.

\section*{Acknowledgment}
The work of Y.O. is supported in part by JSPS Grant-in-Aid for
Scientific Research (B-19340031, S-19104002) and for challenging
Exploratory Research (22656026), and the work of J.Y. is supported
in part by the Air Force Office of Scientific Research (Grant USAF
9550-09-1-0228) and the National Science Foundation (Grant
DMS-0908167).

\section*{Appendix}

\renewcommand{\theequation}{A.\arabic{equation}}

In this appendix we will prove Theorem 1 in section \ref{ratsol}
by using the bilinear method. Firstly we present the following lemma.

\vspace{0.2cm}
\noindent{\bf Lemma 1} \ Let $m_{ij}^{(n)}$, $\varphi_i^{(n)}$ and
$\psi_j^{(n)}$ be functions of $x_1$, $x_2$, $x_{-1}$ and $x_{-2}$
satisfying the following differential and difference relations,
\begin{equation} \label{e:diffrel}
\begin{array}{l}
\partial_{x_1}m_{ij}^{(n)}=\varphi_i^{(n)}\psi_j^{(n)},
\\[5pt]
\partial_{x_2}m_{ij}^{(n)}
=\varphi_i^{(n+1)}\psi_j^{(n)}+\varphi_i^{(n)}\psi_j^{(n-1)},
\\[5pt]
\partial_{x_{-1}}m_{ij}^{(n)}=-\varphi_i^{(n-1)}\psi_j^{(n+1)},
\\[5pt]
\partial_{x_{-2}}m_{ij}^{(n)}
=-\varphi_i^{(n-2)}\psi_j^{(n+1)}-\varphi_i^{(n-1)}\psi_j^{(n+2)},
\\[5pt]
m_{ij}^{(n+1)}=m_{ij}^{(n)}+\varphi_i^{(n)}\psi_j^{(n+1)},
\\[5pt]
\partial_{x_{\nu}}\varphi_i^{(n)}=\varphi_i^{(n+\nu)},
\\[5pt]
\partial_{x_{\nu}}\psi_j^{(n)}=-\psi_j^{(n-\nu)},
\quad
(\nu=1,2,-1,-2).
\end{array}
\end{equation}
Then the determinant,
\begin{equation}
\tau_n=\det_{1\le i,j\le N}\left(m_{ij}^{(n)}\right),
\end{equation}
satisfies the bilinear equations,
\begin{equation} \label{bilin}
\begin{array}{l}
(D_{x_1}D_{x_{-1}}-2)\tau_n\cdot\tau_n=-2\tau_{n+1}\tau_{n-1},
\\[5pt]
(D_{x_1}^2-D_{x_2})\tau_{n+1}\cdot\tau_n=0,
\\[5pt]
(D_{x_{-1}}^2+D_{x_{-2}})\tau_{n+1}\cdot\tau_n=0.
\end{array}
\end{equation}

This lemma can be proved by the same method as for Lemma 3.1 in
\cite{OY}, thus its proof is omitted here.
We note that by the variable transformation
\begin{equation} \label{indepvar}
x_1=\frac{1}{2}(x+y),
\hspace{0.2cm}
x_{-1}=\frac{\epsilon}{2}(x-y),
\hspace{0.2cm}
x_{2}=-\frac{i}{2}t,
\hspace{0.2cm}
x_{-2}=\frac{i}{2}t
\end{equation}
and the complex conjugate condition
\begin{equation} \label{cc}
\overline{\tau}_n=\tau_{-n},
\end{equation}
the above bilinear equations (\ref{bilin}) are reduced to the
bilinear form  (\ref{bilinDSI}) of the DSI equation for $f=\tau_0$ and $g=\tau_1$.
Therefore all we need is to choose appropriate matrix elements $m_{ij}^{(n)}$
which satisfy (\ref{e:diffrel}) and realize the conjugate condition (\ref{cc})
with (\ref{indepvar}).

\vspace{0.2cm}
\noindent \textbf{Proof of Theorem 1} \
It is easy to see that
functions $\varphi_i^{(n)}$, $\psi_j^{(n)}$ and $m_{ij}^{(n)}$
defined by
\begin{equation*}
\varphi_i^{(n)}=p_i^ne^{\xi_i}, \quad
\psi_j^{(n)}=(-q_j)^{-n}e^{\eta_j},
\end{equation*}
\begin{equation*}
m_{ij}^{(n)}=\int^{x_1}\varphi_i^{(n)}\psi_j^{(n)}dx_1
=\frac{1}{p_i+q_j}(-\frac{p_i}{q_j})^ne^{\xi_i+\eta_j},
\end{equation*}
\begin{equation*}
\xi_i=\frac{1}{p_i^2}x_{-2}+\frac{1}{p_i}x_{-1}+p_ix_1+p_i^2x_2,
\end{equation*}
\[
\eta_j=-\frac{1}{q_j^2}x_{-2}+\frac{1}{q_j}x_{-1}+q_jx_1-q_j^2x_2,
\]
satisfy Eqs. (\ref{e:diffrel}). Here $p_i, q_j$ are arbitrary complex constants, and it is assumed that the lower boundary
value of the integral in the above $m_{ij}^{(n)}$ equation is zero.
But these functions do not lead to rational solutions.

To get rational solutions, we differentiate the above functions with
respect to the parameters $p_i$ and $q_j$. To obtain solution expressions in (a) of Theorem 1, we
consider the following
$\varphi_i^{(n)}$, $\psi_j^{(n)}$ and $m_{ij}^{(n)}$ functions:
\begin{equation} \label{e:mijtilde0}
\varphi_i^{(n)}=A_ip_i^ne^{\xi_i}, \quad  \psi_j^{(n)}=B_j(-q_j)^{-n}e^{\eta_j},
\end{equation}
\begin{equation} \label{mijAB0}
m_{ij}^{(n)}=A_iB_j \frac{1}{p_i+q_j}(-\frac{p_i}{q_j})^ne^{\xi_i+\eta_j},
\end{equation}
where $A_i$ and $B_j$ are differential operators of order $n_i$ and
$n_j$ with respect to $p_i$ and $q_j$ respectively, defined as
\begin{equation}
A_i=\sum_{k=0}^{n_i}c_{ik}(p_i\partial_{p_i})^{n_i-k},
\quad
B_j=\sum_{l=0}^{n_j}d_{jl}(q_j\partial_{q_j})^{n_j-l},
\end{equation}
$c_{ik}$, $d_{jl}$ are arbitrary complex constants, and $n_i$ are arbitrary positive integers. It is easy to see that
these functions also satisfy Eqs. (\ref{e:diffrel}), thus
$\tau_n=\det\left(m_{ij}^{(n)}\right)$ with (\ref{mijAB0}) satisfies
the bilinear equations (\ref{bilin}). By using the operator relations
\[(p_i\partial_{p_i})p_i^ne^{\xi_i}=p_i^ne^{\xi_i} (p_i\partial_{p_i}+\xi'_i+n), \]
\[(q_j\partial_{q_j})(-q_j)^{-n}e^{\eta_j}=(-q_j)^{-n}e^{\eta_j} (q_j\partial_{q_j}+\eta'_j-n), \]
where
\[
\xi'_i=-\frac{2}{p_i^2}x_{-2}-\frac{1}{p_i}x_{-1}+p_ix_1+2p_i^2x_2,
\]
\[
\eta'_j=\frac{2}{q_j^2}x_{-2}-\frac{1}{q_j}x_{-1}+q_jx_1-2q_j^2x_2,
\]
the matrix element $m_{ij}^{(n)}$ in (\ref{mijAB0}) becomes
\begin{eqnarray*}
m_{ij}^{(n)}
&=&(-\frac{p_i}{q_j})^ne^{\xi_i+\eta_j}
\sum_{k=0}^{n_i}c_{ik}(p_i\partial_{p_i}+\xi'_i+n)^{n_i-k}  \\
&& \sum_{l=0}^{n_j}d_{jl}(q_j\partial_{q_j}+\eta'_j-n)^{n_j-l}
\frac{1}{p_i+q_j}.
\end{eqnarray*}
Taking parameter constraints
\begin{equation} \label{ccparam}
q_j=\bar p_j,\qquad d_{jl}=\bar c_{jl},
\end{equation}
and using the variable transformation (\ref{indepvar}),
we obtain
\[\eta_j=\bar\xi_j, \quad
\overline{m_{ij}^{(n)}}=  m_{ji}^{(-n)}, \quad
\bar\tau_n=\tau_{-n},
\]
thus the conjugate condition (\ref{cc}) is satisfied. Finally using the gauge freedom of $\tau_n$, we obtain the rational solutions to the DSI equation as given in (a) of Theorem~1.

Next we derive solution expressions in (b) of Theorem 1. For this purpose, we
consider the following
$\varphi_i^{(n)}$, $\psi_j^{(n)}$ and $m_{ij}^{(n)}$ functions:
\begin{equation} \label{e:mijtilde}
\varphi_i^{(n)}=\hat A_ip_i^ne^{\xi_i}, \quad  \psi_j^{(n)}=\hat B_j(-q_j)^{-n}e^{\eta_j},
\end{equation}
\begin{equation} \label{mijAB}
m_{ij}^{(n)}=\int^{x_1}\varphi_i^{(n)}\psi_j^{(n)}dx_1,
\end{equation}
where $\hat A_i$ and $\hat B_j$ are differential operators of order $n_i$ and
$n_j$ with respect to $p_i$ and $q_j$ defined as
\begin{equation}
\hat A_i=\sum_{k=0}^{n_i}\frac{\hat c_{ik}}{(n_i-k)!}(p_i\partial_{p_i})^{n_i-k},
\end{equation}
\begin{equation}
\hat B_j=\sum_{l=0}^{n_j}\frac{\hat d_{jl}}{(n_j-l)!}(q_j\partial_{q_j})^{n_j-l},
\end{equation}
$\hat c_{ik}$, $\hat d_{jl}$ are arbitrary complex constants which are related to the constants $c_{ik}$, $d_{jl}$ in the solution expression (a) as
\[
\hat c_{ik}=c_{ik}(n_i-k)!, \quad \hat d_{jl}=d_{jl}(n_j-l)!,
\]
and it is assumed that the lower boundary value of the integral in (\ref{mijAB}) is zero.
Rewriting $\varphi_i^{(n)}$ and $\psi_j^{(n)}$ in (\ref{e:mijtilde}) as
\begin{eqnarray}
\hat A_ip_i^ne^{\xi_i} & = & P_i^{(n)}p_i^ne^{\xi_i},     \label{Aidef}  \\
\hat B_j(-q_j)^{-n}e^{\eta_j} & = & Q_j^{(n)}(-q_j)^{-n}e^{\eta_j},
\end{eqnarray}
where $P_i^{(n)}$ and $Q_j^{(n)}$ are polynomials of degrees $n_i$ and $n_j$
in $(x_{-2},x_{-1},x_1,x_2)$ respectively, and using the integration by parts, $m_{ij}^{(n)}$ in (\ref{mijAB}) then turns
into
\[
m_{ij}^{(n)}=(-\frac{p_i}{q_j})^ne^{\xi_i+\eta_j}
\sum_{\nu=0}^{n_i+n_j}\frac{(-1)^{\nu}}{(p_i+q_j)^{\nu+1}}
\partial_{x_1}^{\nu}P_i^{(n)}Q_j^{(n)}.
\]
Due to the gauge freedom of $\tau_n$, we can see that
\begin{equation} \label{mijxdiff}
m_{ij}^{(n)}=\sum_{\nu=0}^{n_i+n_j}\left(\frac{-1}{p_i+q_j}\right)^{\nu+1}
\partial_{x_1}^{\nu}P_i^{(n)}Q_j^{(n)}
\end{equation}
gives the same solution. In order for this solution to satisfy the conjugate condition (\ref{cc}), we take
\begin{equation}
q_j=\bar p_j, \quad \hat d_{jl}=\bar{\hat{c}}_{jl}.
\end{equation}
Then in view of the variable transformation (\ref{indepvar}), we have
\begin{equation} \label{appd3}
\overline{Q_j^{(n)}}=P_j^{(-n)}, \quad
\overline{m_{ij}^{(n)}}=  m_{ji}^{(-n)}, \quad
\bar\tau_n=\tau_{-n},
\end{equation}
hence $\tau_n$ with the above matrix elements (\ref{mijxdiff}) satisfies the bilinear DSI equation (\ref{bilinDSI}).

Lastly we derive the explicit expression of $P_i^{(n)}$. From the definition of $P_i^{(n)}$ in (\ref{Aidef}), we have
\begin{eqnarray} \label{appd0}
\sum_{k=0}^{n_i}\frac{\hat c_{ik}}{(n_i-k)!}(p_i\partial_{p_i})^{n_i-k}
p_i^ne^{\xi_i} =P_i^{(n)}p_i^ne^{\xi_i}.
\end{eqnarray}
For $\displaystyle\xi=\sum_{\nu=-2,-1,1,2}p^{\nu}x_{\nu}$, using the functional identity (see Appendix of \cite{OY})
\[e^{\kappa p\partial_p}F(p)=F(e^{\kappa}p), \]
we get
\begin{eqnarray}  \label{appd}
\frac{1}{p^ne^{\xi}}e^{\kappa p\partial_p}p^ne^{\xi}
&=&e^{\kappa n}\exp\Big(\sum_{\nu}(e^{\nu\kappa}-1)p^{\nu}x_{\nu}\Big)  \nonumber \\
&=&\exp\Big(\kappa n+\sum_{k=1}^{\infty}\frac{\kappa^k}{k!}
\sum_{\nu}\nu^kp^{\nu}x_{\nu}\Big)  \nonumber \\
&=&\sum_{k=0}^{\infty}\kappa^kS_k\left(\mbox{\boldmath $\xi$}^{(n)}(p)\right),
\end{eqnarray}
where $\mbox{\boldmath $\xi$}^{(n)}(p)
=(\xi_1(p)+n,\xi_2(p),\cdots,\xi_k(p)+\delta_{k1}n,\cdots)$ and
$\xi_k(p)=\sum_{\nu}\nu^kp^{\nu}x_{\nu}/k!$.
By comparing the coefficient of order $\kappa^k$ in Eq. (\ref{appd}), we obtain
\[
\frac{1}{p^ne^{\xi}}\frac{(p\partial_p)^k}{k!}p^ne^{\xi}
=S_k\left(\mbox{\boldmath $\xi$}^{(n)}(p)\right).
\]
Substituting $p=p_i$ and $\xi=\xi_i$ into this equation and inserting it into (\ref{appd0}), and recalling the variable transformation
(\ref{indepvar}), we then find the explicit expression for
$P_i^{(n)}$ as
\[
P_i^{(n)}=\sum_{k=0}^{n_i}\hat c_{ik}S_{n_i-k}
\left(\mbox{\boldmath $\xi$}^{(n)}(p_i)\right),
\]
where $\mbox{\boldmath $\xi$}^{(n)}(p_i)$ is as given in Theorem 1.
Inserting this $P_i^{(n)}$ and $Q_j^{(n)}$ from
(\ref{appd3}) into (\ref{mijxdiff}), the solution expression in (b) of Theorem 1 is then derived.
This ends the proof of Theorem 1.


\begin{thebibliography}O

\bibitem{rogue_water} C. Kharif, E. Pelinovsky and A. Slunyaev, \emph{Rogue waves in the
ocean} (Springer, Berlin, 2009).

\bibitem{Rogue_nature1} D. R. Solli, C. Ropers, P. Koonath and B. Jalali,
``Optical rogue waves", Nature 450, 1054--1057 (2007).

\bibitem{Rogue_nature2}
B. Kibler, J. Fatome, C. Finot, G. Millot, F. Dias, G. Genty, N.
Akhmediev, J.M. Dudley, ``The Peregrine soliton in nonlinear fibre
optics", Nature Physics, 6, 790--795 (2010).


\bibitem{Peregrine}
D. H. Peregrine, ``Water waves, nonlinear Schr\"odinger equations
and their solutions," J. Australian Math. Soc. B, 25, 16–-43 (1983).

\bibitem{Akhmediev_PRE}
N. Akhmediev, A. Ankiewicz, and J. M. Soto-Crespo, ``Rogue Waves and
Rational Solutions of the Nonlinear Schrödinger Equation," Phys.
Rev. E 80, 026601 (2009).

\bibitem{Rogue_higher_order}
P. Dubard, P. Gaillard, C. Klein, V.B. Matveev, ``On multi-rogue
wave solutions of the NLS equation and positon solutions of the KdV
equation", Eur. Phys. J. Special Topics 185, 247-–258 (2010).

\bibitem{Rogue_higher_order2}
P. Dubard, V.B. Matveev, ``Multi-rogue waves solutions to the
focusing NLS equation and the KP-I equation", Nat. Hazards. Earth.
Syst. Sci. 11, 667--672 (2011).

\bibitem{Rogue_Gaillard}
P. Gaillard, ``Families of quasi-rational solutions of the NLS
equation and multi-rogue waves", J. Phys. A: Math. Theor. 44, 435204
(2011).

\bibitem{Rogue_triplet}
A. Ankiewicz, D.J. Kedziora and N. Akhmediev, ``Rogue wave
triplets", Phys. Lett. A, 375, 2782--2785 (2011).

\bibitem{Rogue_circular}
D.J. Kedziora, A. Ankiewicz, and N. Akhmediev, ``Circular rogue wave
clusters", Phys. Rev. E 84, 056611 (2011).

\bibitem{Liu_qingping}
B. Guo, L. Ling, and Q.P. Liu, ``Nonlinear Schr\"odinger equation:
Generalized Darboux transformation and rogue wave solutions", Phys.
Rev. E 85, 026607 (2012).

\bibitem{OY}
Y. Ohta and J. Yang, ``General high-order rogue waves and their
dynamics in the nonlinear Schr\"odinger equation",
Proc. Roy. Soc. A. 468, 1716–-1740 (2012).

\bibitem{rogue_Hirota}
A. Ankiewicz, J. M. Soto-Crespo, and N. Akhmediev, ``Rogue waves and
rational solutions of the Hirota equation", Phys. Rev. E, 81, 046602
(2010).

\bibitem{Akhmediev_1985}
N. Akhmediev, V.M. Eleonskii, and N.E. Kulagin, ``Generation of a
periodic sequence of picosecond pulses in an optical fiber: Exact
solutions", Sov. Phys. JETP, 62, 894 (1985).

\bibitem{Akhmediev_1988}
N. Akhmediev, V.M. Eleonskii and N.E. Kulagin, ``Exact first-order
solutions of the nonlinear Sch\"odinger equation", Theor. Math.
Phys. 72, 809--818 (1988).

\bibitem{Its_1988}
A.R. Its, A.V. Rybin, and M.A. Salle, ``Exact Integration of
Nonlinear Schr\"odinger equation", Theor. Math. Phys. 74, 29-–45
(1988).

\bibitem{Ablowitz_homo}
M.J. Ablowitz and B.M Herbst, ``On homoclinic structure and
numerically induced chaos for the nonlinear Schr6dinger equation",
SIAM J. Appl. Math. 50, 339 (1990).

\bibitem{Rogue_homo}
N. Akhmediev, A. Ankiewicz and M. Taki, ``Waves that appear from
nowhere and disappear without a trace", Phys. Lett. A 373, 675--678
(2009).

\bibitem{Benney_Roskes}
D. J. Benney and G. Roskes,  ``Wave instabilities", Stud. Appl. Math. 48, 377--385 (1969).

\bibitem{Davey_Stewartson} A. Davey and K. Stewartson,
``On three-dimensional packets of surface waves", Proc. R. Soc. London, A
338, 101--110 (1974).

\bibitem{Ablowitz_book}
M.J. Ablowitz and H. Segur, \emph{Solitons and the Inverse
Scattering Transform} (SIAM, Philadelphia, 1981).

\bibitem{Tajiri_1999}
M. Tajiri and T. Arai, ``Growing-and-decaying mode solution to the
Davey-Stewartson equation", Phys. Rev. E 60, 2297 (1999).

%\bibitem{ACTV}
%M.J. Ablowitz, S. Chakravarty, A.D. Trubatch and J. Villarroel,
%``A novel class of solutions of the non-stationary Schr\"odinger and
%the Kadomtsev-Petviashvili I equations", Phys. Lett. A 267, 132--146, 2000.

\end{thebibliography}
\end{document}